\documentclass[letterpaper, 10 pt, conference]{ieeeconf}
\IEEEoverridecommandlockouts
\usepackage{cite}
\usepackage{amsmath,amssymb,amsfonts}
\usepackage{hyperref}
\usepackage{algorithmic}
\usepackage{graphicx}
\usepackage{textcomp}
\usepackage{xcolor}
\usepackage{multirow}
\usepackage{tabularx}
\usepackage{booktabs}
\usepackage{xurl}
\def\BibTeX{{\rm B\kern-.05em{\sc i\kern-.025em b}\kern-.08em
    T\kern-.1667em\lower.7ex\hbox{E}\kern-.125emX}}
\begin{document}

\title{Towards Low-Barrier Cybersecurity Research and Education for Industrial Control Systems

\author{Colman McGuan, Chansu Yu,  Qin Lin% <-this % stops a space
%\thanks{*This work was not supported by any organization}% <-this % stops a space
\thanks{$^{}$All authors are with the Department of Electrical Engineering and Computer Science, Cleveland State University, 2121 Euclid Avenue, Cleveland, OH 44115, USA.
Corresponding author: Qin Lin, {\tt\small q.lin80@csuohio.edu}}%
}
}

% \author{\IEEEauthorblockN{Colman McGuan}
% \IEEEauthorblockA{\textit{Department of EECS} \\
% \textit{Cleveland State University}\\
% Cleveland, USA \\
% c.mcguan@vikes.csuohio.edu}
% \and
% \IEEEauthorblockN{Chansu Yu}
% \IEEEauthorblockA{\textit{Department of EECS} \\
% \textit{Cleveland State University}\\
% Cleveland, USA \\
% c.yu91@csuohio.edu}
% \and
% \IEEEauthorblockN{Qin Lin}
% \IEEEauthorblockA{\textit{Department of EECS} \\
% \textit{Cleveland State University}\\
% Cleveland, USA \\
% q.lin80@csuohio.edu}
% }

%\author{Colman McGuan, Chansu Yu, Qin Lin % <-this % stops a space
% %\thanks{*This work was not supported by any organization}% <-this % stops a space
% \thanks{$^{}$All authors are with the Department of Electrical Engineering and Computer Science, Cleveland State University, 2121 Euclid Avenue, Cleveland, OH 44115, USA. Corresponding author: Qin Lin, {\tt\small q.lin80@csuohio.edu}}%
% }

%\and
%\IEEEauthorblockN{4\textsuperscript{th} Given Name Surname}
%\IEEEauthorblockA{\textit{dept. name of organization (of Aff.)} \\
%\textit{name of organization (of Aff.)}\\
%City, Country \\
%email address or ORCID}
%\and
%\IEEEauthorblockN{5\textsuperscript{th} Given Name Surname}
%\IEEEauthorblockA{\textit{dept. name of organization (of Aff.)} \\
%\textit{name of organization (of Aff.)}\\
%City, Country \\
%email address or ORCID}
%\and
%\IEEEauthorblockN{6\textsuperscript{th} Given Name Surname}
%\IEEEauthorblockA{\textit{dept. name of organization (of Aff.)} \\
%\textit{name of organization (of Aff.)}\\
%City, Country \\
%email address or ORCID}
%}

\maketitle

\begin{abstract}
The protection of Industrial Control Systems (ICS) that are employed in public critical infrastructures is of utmost importance due to catastrophic physical damages cyberattacks may cause. The research community requires testbeds for validation and comparing various intrusion detection algorithms to protect ICS. However, there exist high barriers to entry for research and education in the ICS cybersecurity domain due to expensive hardware, software, and inherent dangers of manipulating real-world systems. To close the gap, built upon recently developed 3D high-fidelity simulators, we further showcase our integrated framework to automatically launch cyberattacks, collect data, train machine learning models, and evaluate for practical chemical and manufacturing processes. On our testbed, we validate our proposed intrusion detection model called Minimal Threshold and Window SVM (MinTWin SVM) that utilizes unsupervised machine learning via a one-class SVM in combination with a sliding window and classification threshold. Results show that MinTWin SVM minimizes false positives and is responsive to physical process anomalies. Furthermore, we incorporate our framework with ICS cybersecurity education by using our dataset in an undergraduate machine learning course where students gain hands-on experience in practicing machine learning theory with a practical ICS dataset. All of our implementations have been open-sourced.
\end{abstract}


\section{Introduction}
Industrial Control Systems (ICS) are used to electronically automate the control of an industrial process. Traditionally, ICS were not networked with the outside world. However, the onset of Industry 4.0 has increased the number of industrial components connected to the Internet and has therefore increased their vulnerability to cyberattacks \cite{ref:grfics,ref:rsa,ref:swat,ref:injtool,ref:rnn,ref:factoryio}.

We focus on intrusion detection based on physical process anomalies. The underlying concept of detecting physical process anomalies caused by cyberattacks is to compare observed behaviors with the expected behavior based on physical invariants, which are properties of a physical system that always hold true due to immutable physical laws. Data-centric approaches are gaining popularity as a powerful tool to automatically discover invariants from an ICS's normal operational data without expert knowledge, e.g., Short-Term Memory Recurrent Neural Network (LSTM-RNN) \cite{ref:rnn}, explainable graphic model \cite{lin2018tabor}, residual skewness analysis \cite{ref:rsa}, etc.

In recent years, physical testbeds have been made available such as SWaT \cite{ref:swat} (water treatment system), WADI \cite{ahmed2017wadi} (water distribution system), EPIC \cite{adepu2019epic} (electric power system), etc.
However, the entry point for designing and evaluating detection models for ICS is costly due to expensive hardware, software, and inherent dangers of manipulating real-world systems. 

To overcome these obstacles, we turn to two simulated platforms: GRFICSv2~\cite{ref:grfics} and Factory I/O~\cite{ref:factoryio}. Both platforms simulate real-world ICS, including network traffic and physical dynamics. Both also provide real-time graphics to visualize the setup as well as view the effects of attacks. The use of an entirely simulated platform with virtual machines (VMs) allows us to conveniently modify the setup for automated attack launching and data collection.

We introduce a novel intrusion detection model for deployment as an online monitor in ICS called Minimal Threshold and Window SVM (MinTWin SVM). We design our model around avoiding the costliness of false positives and utilize a one-class SVM with a sliding window and predefined threshold for classification where the threshold determines the percentage of anomaly data points required in the window before an attack is identified. 

Lastly, to assist in developing a globally competitive cybersecurity workforce in our school's education program, we tightly integrate our framework with a project-centric machine learning course for undergraduates \cite{lin2020using}. Our students have the opportunity to practice machine learning models such as Gaussian analysis on high-quality datasets from practical ICS.

The contributions we make in this work are as follows:

\begin{enumerate}
    \item Built upon realistic/graphic simulated testbeds---namely, GRFICSv2 and Factory I/O---we further design an open-source, realistic, and comprehensive framework to include automated attack launching, dataset generation, and evaluation. The framework is available to the research community for validation, comparison, and benchmarking of various anomaly detection algorithms;
    \item We design a novel intrusion detection model named MinTWin SVM and demonstrate its ability to detect a variety of attacks as well as maintain accuracy in different ICS settings. Furthermore, we provide evidence that MinTWin SVM is suitable for real-world deployment for continuous ICS monitoring;
    \item We integrate our research with cybersecurity and ICS education: the GRFICSv2 dataset has been used in the mid-term project of our undergraduate machine learning course. Students gain hands-on experience in practicing machine learning theory with a practical ICS dataset.
\end{enumerate}

The rest of this paper is organized as follows. Our attack categories are defined in Sec. \ref{sec:attack-categories}. We elaborate on the proposed MinTWin SVM in Sec. \ref{sec:mintwin}. The setup of our framework on GRFICSv2 and Factory I/O is introduced in Sec. \ref{sec:exp-grfics} and Sec. \ref{sec:exp-factoryio}, respectively. The experimental results are presented in Sec. \ref{sec:experiment}. The integration of education is discussed in Sec. \ref{sec:education}. Concluding remarks and future work are given in Sec. \ref{sec:conclusion}.

\section{Attack Categories}
\label{sec:attack-categories}
We categorize our attacks to clearly define which parts of the ICS are under attack. We split our attacks into five distinct categories with a sixth category that overlaps the others. %In general, attacks begin simple and gradually increase in complexity as they consider more components of the ICS. The definition of each attack category is given below.

\subsubsection{Single Sensor Attacks} The value of a single sensor reading from any point in the ICS is intercepted and manipulated. %For example, if a binary discrete sensor reads as on, it is flipped to off.

\subsubsection{Single Actuator Attacks} %In \emph{Single Actuator Attacks}, 
The value the PLC sends to an actuator is intercepted and manipulated. %For example, suppose the PLC sends a Modbus packet instructing a valve to close; after intercepting the packet, we spoof the value to instruct the valve to remain open.

\subsubsection{Multiple Sensor Attacks} %In \emph{Multiple Sensor Attacks}, the value of multiple sensor readings in the ICS are intercepted and manipulated. 
The values of at least two distinct sensors are manipulated.

\subsubsection{Multiple Actuator Attacks} %In \emph{Multiple Actuator Attacks}, the values of two or more actuators are intercepted in transit from the PLC to ICS subcomponents and manipulated.
The values sent by the PLC to at least two distinct actuators are manipulated.

\subsubsection{Complex Attacks} In \emph{Complex Attacks}, any combination of both sensors and actuators are manipulated. However, it is required that at least one sensor and at least one actuator is manipulated---that is, \textit{Complex Attacks} cannot consist of only sensor manipulation, nor can they consist of only actuator manipulation.

\subsubsection{Stealthy Attacks} It is challenging to detect \emph{Stealthy Attacks} due to their somewhat benign implementation method that mimics or barely modifies normal plant operation \cite{ref:rsa}. We implement these to test the resilience of our model against harder-to-detect attacks. An example is intercepting the reading of a continuous sensor and increasing or decreasing the value by 10\% of the total sensor range. This is harder to detect because although the value has been spoofed, it does not deviate extremely from the expected value. \textit{Stealthy Attacks} are not a distinct category but are rather distributed among the other categories.

\section{MinTWin SVM Intrusion Detector}
\label{sec:mintwin}
When designing our intrusion detection model, we bear three goals in mind: 1) using unsupervised machine learning, because labeling data is particularly difficult for ICS security---typically, there is access to a large amount of benign data while malicious data is rare;  2) minimizing the feature set and amount of data required for a scalable implementation; 3) minimizing false positives, because erroneously identifying a system as under attack and preemptively shutting it down can have serious consequences, especially for safety-critical systems.

\subsection{Feature Set}
As a physical process anomaly detector, the training data and features of MinTWin SVM are sourced from the physical process, i.e., any sensor and actuator values, which we collect from the PLC. We introduce another feature by calculating the difference in sensor values from our previous reading to capture the dynamic change of information. We also include the outputs from the PLC to aid in identifying discrepancies between a read actuator value (actual command) and a desired command.

\subsection{Constructing MinTWin SVM}
To construct and fine-tune MinTWin SVM, we use the GRFICSv2 attack and benign datasets outlined below. First, we explain our rationale for the choice of the classifier.

\subsubsection{Choosing a Classifier}  %Unsupervised machine learning allows us to train the classifier using only benign data and consider anything outside to be abnormal.
We compare and test existing successful anomaly detection models in the literature: \emph{One Class SVM} \cite{scholkopf2001estimating}, \emph{Isolation Forest} \cite{liu2008isolation}, and \emph{Local Outlier Factor} \cite{breunig2000lof}. We incorporate a \emph{One Class SVM} into our model. A comparison with other models is shown in the experimental results in Sec. \ref{sec:experiment}.

\subsubsection{OneClassSVM Hyperparameters}
The choice of hyperparameter values is outlined below.

%We do extensive testing by tuning the available parameters to create a high-performing model. 
\textbf{Tuning the Contamination Threshold:} The $\nu$ parameter importantly determines the maximum percentage of training data that can be misclassified as anomalies \cite{scholkopf2001estimating}. After testing, we arrive at a value of 0.05 for our model. %This value is small enough such that it is not too sensitive to slight deviations in data but also does not necessarily overfit our training data.

\textbf{Choosing a Kernel:} We have tested all commonly used kernels to find the best-performing option, such as \textit{linear}, \textit{poly}, \textit{RBF}, \textit{sigmoid}, and \textit{precomputed}. We find that the \textit{RBF} kernel is best suited for our model as it identifies the most attacks and generates the fewest number of false positives on benign data.

\subsubsection{Choosing a Sliding Window Size and Classification Threshold} We employ a sliding window methodology in our model that functions exactly like a queue. To find the optimal window size, we test with values of 5, 10, 15, and 20 seconds.

We must also select an optimal threshold for classification---that is, the percentage of data points within the window that must be classified as abnormal before the system is considered to be under attack. We select three thresholds of 60\%, 70\%, and 80\% for testing.

To find the optimal pairing of sliding window size and classification threshold, each window size is tested with each threshold value. We calculate the true positive rate (TPR) and false positive rate (FPR). After comparing the performance of each configuration, we settle on a 15-second window with a 60\% threshold. As such, we name the model Minimal Threshold and Window SVM (MinTWin SVM).

We find that the performance of MinTWin SVM is tied to the length of the control cycle---the sensor or actuator whose readings take the longest to repeat under normal operation---of the ICS. We observe this value to be roughly 1,000 seconds on GRFICSv2 and 400 seconds on Factory I/O. This relationship is discussed further below.

\section{Experimental Setup on the GRFICSv2 Platform}
\label{sec:exp-grfics}
Each of our chosen simulation platforms uses VirtualBox \cite{ref:vbox} to emulate various ICS components. We use VBoxManage \cite{ref:vboxmanage} to remotely administer each VM. This allows us to start, stop, and execute programs on a VM from the host machine. We incorporate these actions into a simple shell script that enumerates through each attack and benign data capture. We can therefore automate attack launching and data collection such that it does not require continually monitoring the setup and manually performing the experiment.

We make the code for our setup on both GRFICSv2\footnote{GRFICSv2 setup: \url{https://github.com/csmcguan/mintwin-svm-grfics}} and Factory I/O\footnote{Factory I/O setup: \url{https://github.com/csmcguan/mintwin-svm-factoryio}} available to enable future research to replicate our experiment and expand upon our work %\footnote{\href{https://drive.google.com/drive/folders/1syutctNYnZnK6IoxQh-Rdbxvkq_VpgAY?usp=share_link}{This link is colored blue.}}.

\subsection{GRFICSv2 Overview}
GRFICSv2 is a simulation with 3D visualization based on \cite{ref:chemctl} and outlined in detail in \cite{ref:grfics}. The simulation emulates a two-phase chemical plant process that seeks to maximize the efficiency and yield of the reaction and minimize the amount of each chemical purged  \cite{ref:grfics,ref:chemctl}. Within the simulation, there are four actuators (valves) and nine sensors.

The simulation consists of five VMs that emulate plant operations and real network traffic; there is a simulation backend VM, a PLC VM running OpenPLC \cite{ref:openplc}, a pfSense VM, a Human Machine Interface (HMI) VM, and a workstation VM. Communication between the PLC, HMI, and simulation uses Modbus TCP/IP---the chemical plant simulation VM implements Modbus servers that listen on virtual IP addresses that correspond to different plant components \cite{ref:grfics}. The use of Modbus TCP/IP allows us to design and perform man-in-the-middle (MITM) attacks on sensor and actuator values by spoofing the values in Modbus packets between the PLC and chemical plant components.

\subsubsection{Chemical Plant Simulation Components} 
\begin{figure}[htbp]
    \centering
    \includegraphics[width=0.7\linewidth]{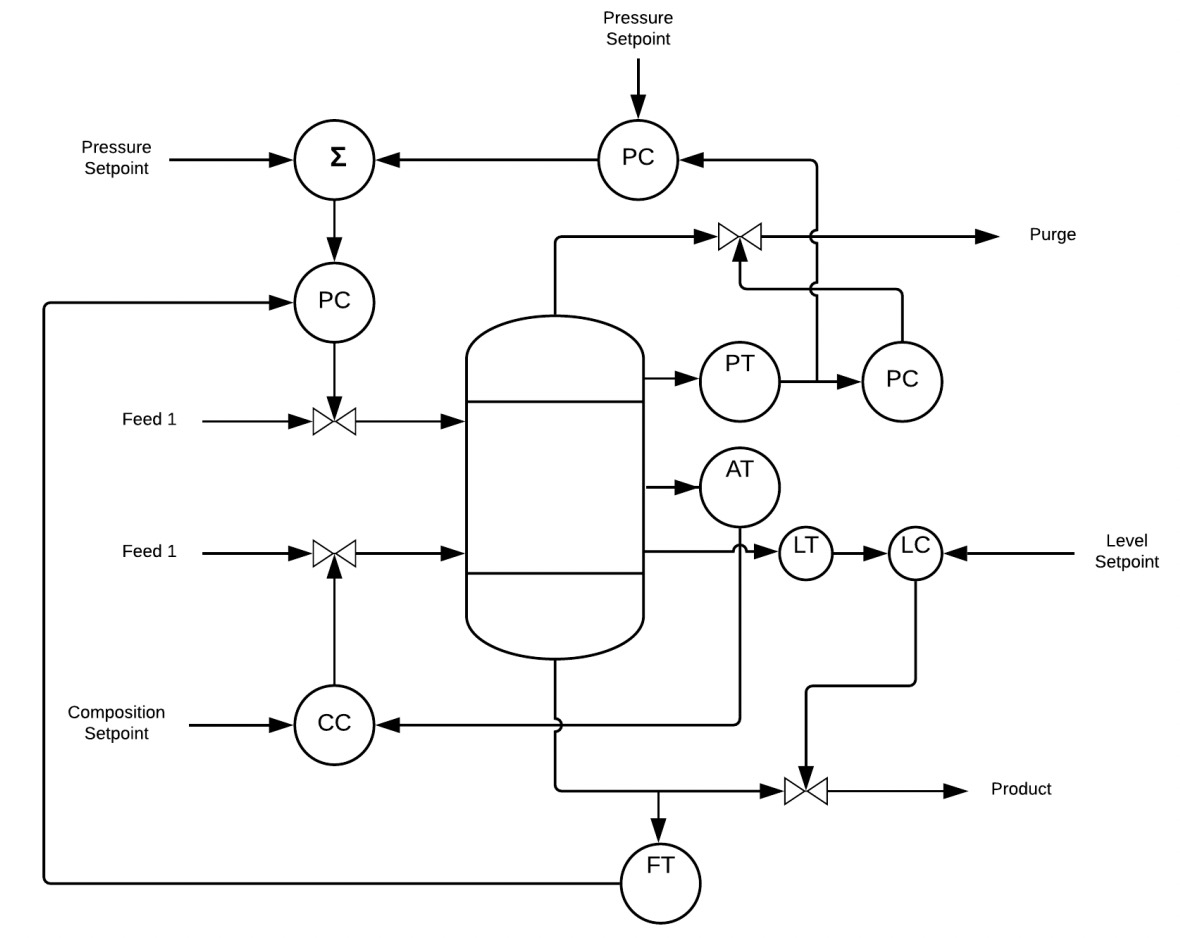}
    \caption{Piping \& Instrumentation Diagram (P\&ID) for
the chemical plant of the Tennessee Eastman Challenge Process.} 
    \label{fig:chemical-plant}
\end{figure}

All sensors and actuators have a possible range from 0 to 65,535. As shown in Fig. \ref{fig:chemical-plant}, the chemical plant consists of five sub-components: the Tank, Feed 1, Feed 2, Product, Purge, and Composition. The Tank takes input from Feed 1 and Feed 2, and yields output to Product and Purge. The Product is the main outflow of the chemical plant whereas the Purge acts as a pressure override. Each input and output is equipped with a valve that controls the flow rate through it. Finally, the Composition indicates the percentage of each chemical in the Purge.

\subsection{Virtual Machine Modifications}
We incorporate data collection and the launching of attacks into the existing VM topology to simulate a real-world scenario. Our setup assumes that the workstation VM has been compromised by a malicious actor who will launch MITM attacks between the PLC and chemical plant components. Data collection takes place on the HMI VM to mimic a real-world setup.

\subsubsection{Workstation VM Modifications} We use Ettercap \cite{ref:ettercap} on the workstation VM to carry out MITM attacks. Ettercap enables us to ARP poison the PLC VM and chemical plant components to intercept their communication. Using Ettercap’s built-in etterfilter \cite{ref:etterfilter} feature, we spoof sensor and actuator values in Modbus packets as desired.

\subsubsection{HMI VM Modifications} In the GRFICSv2 platform, the HMI VM is on a separate network from the other VMs, with the pfSense VM acting as a firewall between the two networks. This makes the HMI VM an ideal candidate for data collection to take place due to its isolation from other components.

Recall that our feature set is strictly limited to data available from the PLC. We use PyModbus \cite{ref:pymodbus} to poll the necessary data from the PLC. We set up a ModbusTcpClient \cite{ref:pymodbus} and log features to a CSV file.

% \subsection{Control Cycles and Stabilizing the Chemical Plant}
% It is crucial that we find the point at which the chemical plant stabilizes from its initial boot. This serves two purposes: first, our training data should reflect normal plant operations and not contain data with fluctuations as the reaction stabilizes, and second, to launch attacks from a basepoint of normal behavior to ensure that the attack dataset contains data strictly as a result of launching attacks and none due to chemical plant initialization.

% To find the stabilization point, we start the chemical plant simulation and plot each sensor and actuator. Most components stabilize relatively quickly except for Feed 2, which we find takes around 5,000 seconds (roughly 83 minutes) to stabilize. We round this up to 90 minutes and use it as our baseline for data collection. We take a snapshot of each VM after 90 minutes so that each data capture is guaranteed to be initiated from the same state.

% We also must identify the control cycle length of the chemical plant---that is, each sensor and actuator’s values appear as a cycle of values when visualized. We take the longest cycle and use it for our data collection period. Feed 2 has the longest cycle at around 1,000 seconds, which is therefore the value we use for our data collection period.

\section{Experimental Setup on the Factory I/O Platform}
\label{sec:exp-factoryio}
\subsection{Factory I/O Overview}
We implement the Production Line scene \cite{ref:prodline} using the control logic provided by \cite{ref:prodlinectl}. The scene consists of two robotic arms that take raw material from conveyor belts and place it into two machining centers. Once finished, the robotic arms take the pieces out of the machining centers and put them on a separate conveyor belt where they are carried to the end of the assembly line. In this setup, our feature set consists of 12 sensors and 15 actuators. The majority of the sensors and  actuators are binary discrete values---i.e., 1 or 0---except for two of the sensors, which are continuous, and therefore must be normalized. The control logic for this setup is likewise implemented using OpenPLC and Modbus TCP/IP. This allows us to perform spoofing attacks on the system between the PLC and Production Line simulation.

\subsection{Factory I/O Setup Virtual Machines}
We implement three VMs in our setup. We run a vanilla installation of Debian GNU/Linux on each.

\subsubsection{PLC VM} The PLC VM implements the required control logic for the Production Line scene. It runs OpenPLC and communicates with the simulation using Modbus TCP/IP.

\subsubsection{HMI VM} The HMI VM plays the role of data collection. We use a ModbusTcpClient from PyModbus to poll data from the PLC and log it to a CSV file.

\subsubsection{Attacker VM} The Attacker VM launches MITM attacks between the PLC and simulation by spoofing the Modbus packets.

\subsection{Caveats}
We cannot use Ettercap because it only allows for byte-wise comparison and modification of network packets. This is not an issue on the GRFICSv2 platform because each sensor and actuator are a 2-byte value. In contrast, the binary discrete sensors and actuators in the Factory I/O setup require bitwise comparison and modification. Therefore, we create our own tool to ARP poison the Factory I/O simulation and PLC to perform MITM attacks. We use NetfilterQueue \cite{ref:nfq} and Scapy \cite{ref:scapy} to intercept and modify Modbus packets in real time.

The size of the sliding window must also be adjusted to correspond to the reduced control cycle time. We adjust the sliding window size by starting with the same ratio of the sliding window size to the control cycle length of GRFICSv2 and doing any necessary fine-tuning. We find that a sliding window size of five seconds yields the best results.

% \subsection{Adjusting the Sliding Window Size}
% The control cycle length in the Production Line scene on the Factory I/O platform is less than half that of the GRFICSv2 platform. It is therefore necessary to modify the size of the sliding window used in our model; otherwise, it may be too large to capture smaller disturbances and too slow to be practical. We adjust the sliding window size by starting with the same ratio of the sliding window size to the control cycle length of GRFICSv2---i.e., 15 seconds is 1.5\% of the 1,000-second control cycle length---and doing any necessary fine-tuning. We settle on a sliding window size of five seconds.

\section{Experimental Results}
\label{sec:experiment}
\subsection{Datasets}
Here, we outline our datasets. Note that each data capture is launched for an entire control cycle.
\subsubsection{GRFICSv2 Attack Dataset}
We create 54 attacks. The number of attacks belonging to each category is given below in Table~\ref{tab1}.

\begin{table}[htbp]
    \centering
    \caption{Number of Attacks by Category on GRFICSv2.}
    \begin{tabularx}{\linewidth}{
        >{\centering\arraybackslash}X
        >{\centering\arraybackslash}X }
        \toprule
        Attack Category & Number of Attacks \\
        \midrule
        Single Sensor & 28 \\
        Single Actuator & 8 \\
        Multiple Sensor & 10 \\
        Multiple Actuator & 6 \\
        Complex & 2\\
        Stealthy & 16\\
        \bottomrule
    \end{tabularx}
    \label{tab1}
\end{table}

\subsubsection{Factory I/O Attack Dataset}
Given the large attack dataset on the GRFICSv2 platform, the attack dataset on the Factory I/O platform is considerably smaller. Here, we implement seven attacks, two of which are \textit{Multiple Sensor Attacks} and five of which are \textit{Complex Attacks}. As each machining center functions independently, we include attacks that target one machining center and attacks that target both centers simultaneously.

\subsubsection{Benign Datasets}
We use benign data to verify that MinTWin SVM is not overly sensetive such that it misclassifies benign data. For each platform, we collect 50 successive captures---that is, we allow the simulation to run continuously and collect 50 benign captures in a row. This verifies that our model does not lose its resilience as the data being collected drifts further from our initial starting point. The 50 captures of benign data equate to nearly 14 hours of continuous plant operation on GRFICSv2 and 5.5 hours on Factory I/O.

\subsection{Results}
% In this section, we evaluate MinTWin SVM using the two metrics outlined in Section 3.3. First, we look at the effectiveness against the pre-collected attack and benign datasets. Then, we explore the detection time required to evaluate our model's responsiveness.

\subsubsection{MinTWin SVM Evaluation on Pre-Collected Datasets}
\begin{figure}[htbp]
    \centering
    \includegraphics[width=0.8\linewidth]{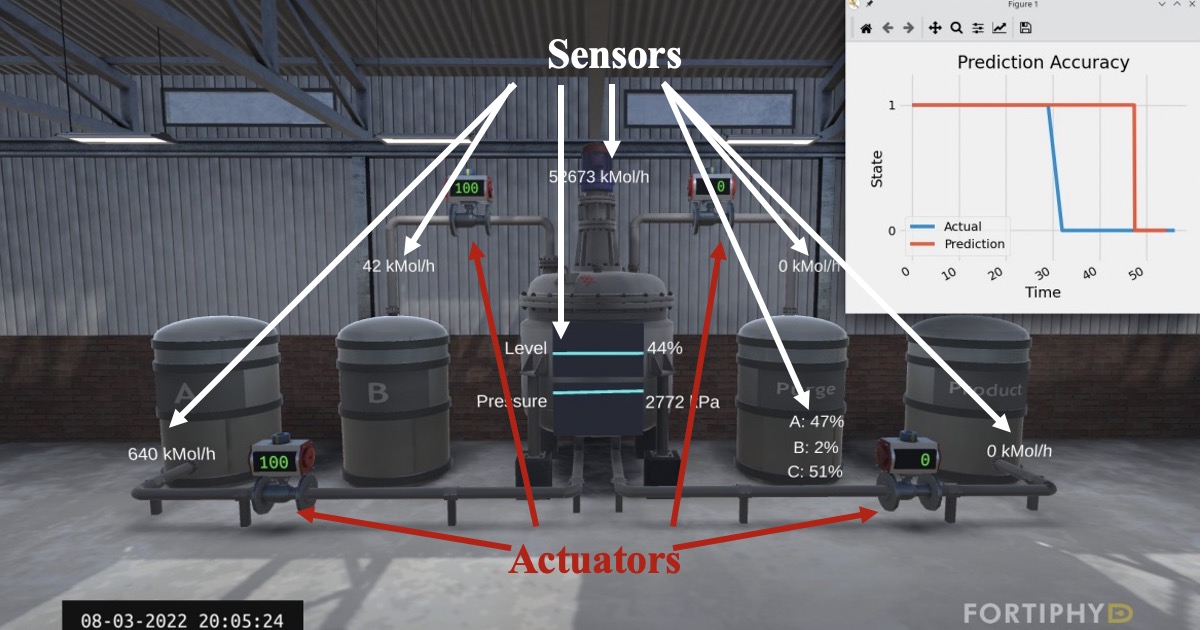}
    \caption{Screenshot of Our GRFICSv2 Demo} 
    \label{fig:GRFICS-demo}
\end{figure}

\begin{figure}[htbp]
    \centering
    \includegraphics[width=0.8\linewidth]{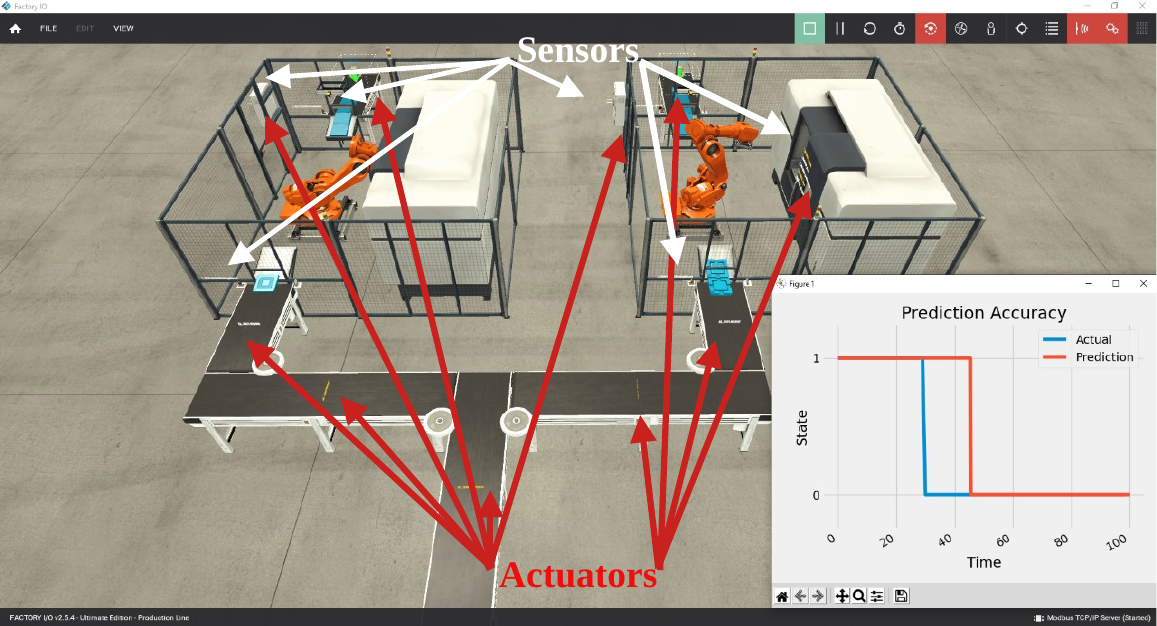}
    \caption{Screenshot of Our Factory I/O Demo} 
    \label{fig:factoryio-demo}
\end{figure}

Here, we test MinTWin SVM against the pre-collected attack and benign datasets. The number of true positives (TPs), false positives (FPs), true negatives (TNs), and false negatives (FNs) are shown in Table~\ref{tab2}. Figs. \ref{fig:GRFICS-demo} and \ref{fig:factoryio-demo} show screenshots of our demos on GRFICSv2 and Factory I/O respectively. The demos can also be found online\footnote{GRFICSv2 demo: \url{https://youtu.be/ckglvMokx6M}}\textsuperscript{,}\footnote{Factory I/O demo: \url{https://youtu.be/h-0m85NjdCc}}.

\begin{table}[htbp]
%\small
    \centering
    \caption{Classification Results Against Pre-Collected Datasets.}
    \begin{tabularx}{\linewidth}{
    >{\centering\arraybackslash}X
    >{\centering\arraybackslash}X
    >{\centering\arraybackslash}X}
        \toprule
        %\midrule
        Platform & Metric & Quantity \\
        %\midrule
        \midrule
        \multirow{4}{*}{GRFICSv2} & TP & 48 \\
        & FP & 0 \\
        & TN & 50 \\
        & FN & 6 \\
        \midrule
        \multirow{4}{*}{Factory I/O} & TP & 7 \\
        & FP & 0 \\
        & TN & 50 \\
        & FN & 0 \\
        %\midrule
        \bottomrule
    \end{tabularx}
    \label{tab2}
\end{table}

MinTWin SVM generates no false positives while still identifying most attacks on GRFICSv2 and \textit{all} attacks on Factory I/O, with a TPR of 88.9\% and 100.0\%, respectively. This indicates that our model is adaptable to different industrial settings and is not deterred by a change in the sensor and actuator type---that is, it is accurate on both discrete and continuous values. The only requirement to transfer MinTWin SVM to a different industrial setting is to adjust the size of the sliding window to correspond to the control cycle length.

For further comparison, we evaluate MinTWin SVM against approaches using the \textit{Isolation Forest} and \textit{Local Outlier Factor} classifiers discussed above. For each classifier, we use the same sliding window size and classification threshold as in MinTWin SVM. The results are given in Table~\ref{tab3}. Across both platforms, MinTWin SVM produces the best results. The \textit{Local Outlier Factor} is promising on the Factory I/O datasets, but fails to replicate those results on the GRFICSv2 datasets. This may be explained by the difference in sensor types---i.e., continuous vs. discrete---between the two platforms. The \textit{Isolation Forest} classifier is more stable than the \textit{Local Outlier Factor}, but has a lower TPR than MinTWin SVM.

\begin{table}[htbp]
    \centering
    \caption{TPR and FPR of Classifiers Against Pre-Collected Datasets.}
    \begin{tabularx}{\linewidth}{
    >{\centering\arraybackslash}c
    >{\centering\arraybackslash}X
    >{\centering\arraybackslash}c
    >{\centering\arraybackslash}c}
        \toprule
        %\midrule
        Platform & Classifier & TPR & FPR \\
        %\midrule
        \midrule
        \multirow{3}{*}{GRFICSv2} & MinTWin SVM & \textbf{88.9\%} & \textbf{0.0\%} \\
        & Isolation Forest \cite{liu2008isolation} & 18.4\% & 0.0\% \\
        & Local Outlier Factor \cite{breunig2000lof} & 100.0\% & 100.0\% \\
        \midrule
        \multirow{3}{*}{Factory I/O} & MinTWin SVM & \textbf{100.0\%} & \textbf{0.0\%} \\
        & Isolation Forest \cite{liu2008isolation} & 85.7\% & 0.0\% \\
        & Local Outlier Factor \cite{breunig2000lof} & \textbf{100.0\%} & \textbf{0.0\%} \\
        %\midrule
        \bottomrule
    \end{tabularx}
    \label{tab3}
\end{table}

\subsubsection{Attack Identification by Category on GRFICSv2} 
We show the breakdown of attack identification by category on GRFICSv2. That is, we calculate the percentage of attacks correctly identified in each of our previously defined categories. The results are given in Table~\ref{tab4}.

\begin{table}[htbp]
    \centering
    \caption{Attack Identification by Category.}
    \begin{tabularx}{\linewidth}{
    >{\centering\arraybackslash}X
    >{\centering\arraybackslash}X}
        %\midrule
        \toprule
        Attack Category & Percentage Identified \\
        %\midrule
        \midrule
        Single Sensor & 78.6\% \\
        Single Actuator & 100.0\% \\
        Multiple Sensor & 100.0\% \\
        Multiple Actuator & 100.0\% \\
        Complex & 100.0\% \\
        Stealthy & 62.5\% \\
        %\midrule
        \bottomrule
    \end{tabularx}
    \label{tab4}
\end{table}

MinTWin SVM identifies 100\% of attacks in four of five official categories. All six unidentified attacks belong to \textit{Stealthy Attacks} and all of them belong to the \textit{Single Sensor} category. Recalling that \textit{Stealthy Attacks} are by design difficult to detect and include behavior that is somewhat benign, the fact that MinTWin SVM identifies 62.5\% (or 10 of 16) of \textit{Stealthy Attacks} points to its resilience and successful deployment.

Although a configuration could have been chosen which identifies more \textit{Stealthy Attacks}---e.g., a 5-second window with a 60\% classification threshold identifies 93.8\% (15 of 16) of \textit{Stealthy Attacks}---this increase in accuracy comes at the cost of a much more pronounced increase in FP. In fact, the FPR increases to 82.0\% for such a configuration as opposed to 0\% in MinTWin SVM. Given our previously stated concern for FPs, identifying five more stealthy attacks is not worth the trade-off of an increase in the TPR to 82.0\%.

\subsubsection{Attack Detection Time}
%In this section, we analyze the amount of time it takes MinTWin SVM to identify each attack. 

% \textbf{Attack Detection Time Experimental Setup:} We create a simple setup that launches each attack for a full control cycle. We start MinTWin SVM at this point, which is configured to predict a safe state until the sliding window is full because there is not yet enough data to make a decision. We launch the attack at time 30 seconds. As soon as the attack is identified, it writes the time in seconds since the attack was launched to the log and continues to the next one. If an attack was unable to be identified in under one control cycle, the attack is stopped and a value of -1 is written to the log. For example, as shown in Fig. \ref{fig:GRFICS-demo}, an attack happens 30 seconds from the beginning, we detect it at around 50 seconds. Therefore, the detection time is around 20 seconds in this scenario.

For each identified attack, we calculate the median detection time---that is, we launch each attack and log the time taken to identify it in real time. The results are given in Table~\ref{tab5}. The results demonstrate that MinTWin SVM identifies attacks quickly.

\begin{table}[htbp]
    \centering
    \caption{Median Attack Detection Time.}
    \begin{tabularx}{\linewidth}{
    >{\centering\arraybackslash}X
    >{\centering\arraybackslash}X}
        \toprule
        %\midrule
        Platform & Time (seconds) \\
        \midrule
        %\midrule
        %\multirow{2}{*}{GRFICSv2} & Mean Detection Time & 62.5 \\
        GRFICSv2 & 17.7 \\
        %\multirow{2}{*}{} & Mean Detection Time & 11.2 \\
        Factory I/O & 8.9 \\
        %\midrule
        \bottomrule
    \end{tabularx}
    \label{tab5}
\end{table}

% Notice that on GRFICSv2 the median detection time is nearly half of the mean detection time. This can be explained by attacks 7 and 45, which have a detection time of 914.0 and 778.9 seconds respectively. When we remove these data points from the calculation, the mean detection time drops to 31.49 seconds. This is more in line with the speed of MinTWin SVM as given by the median detection time; in fact, of the identified attacks on GRFICSv2, 62.5\% have a detection time under 20 seconds and 81.3\% under 30 seconds. This highlights the responsiveness of our model in identifying attacks.

% The detection time on Factory I/O is smaller than on GRFICSv2. This is likely owed to the decrease in the sliding window size. This also indicates that across different platforms, our model maintains its responsiveness. On Factory I/O, we observe one attack with a detection time of 33.2 seconds. When we remove this data point from the calculation of the mean detection time, it falls to 6.11 seconds. Furthermore, 71.4\% (5 out of 7) of attacks have a detection time below 10 seconds. 

\section{Integration with Education}
\label{sec:education}

The education goal of our framework is to support the development of a diverse, globally competitive cybersecurity workforce in our education program. We aim to lower the entry barriers to ICS cybersecurity, and improve accessibility for students to learn principles and defense of ICS. \emph{Introduction to Machine Learning} is an elective course at our university for undergraduate students majoring in electrical engineering and computer science. The course is project-centric. After learning machine learning theory in class, students have opportunities to practice on real-world datasets to obtain hands-on experience. Applying machine learning techniques in the cybersecurity domain is gaining popularity in recent years. However, unlike other domains, high-quality cybersecurity datasets---especially for ICS---are rare due to the high cost of platforms and the dangers of manipulating real-world systems. Our framework built upon simulated testbeds perfectly fulfills the requirement.

The project was given to students as the mid-term exam. We had 12 teams formed by 22 students; each team had no more than two members. Our GRFICSv2 dataset was split into three parts: training (1 CSV file), validation (23 CSV files), and testing (58 CSV files). Each CSV file can be benign (label 0) or malicious (label 1). The students were told that the training dataset is benign. In the validation dataset, they were given labels for each scenario, i.e., each CSV file. Using these labels, students could tune the hyperparameters of their machine learning models. In our lecture, we taught anomaly detection using \emph{Independent Gaussian Analysis} (IGA) assuming all features are independent and \emph{Multivariate Gaussian Analysis} (MGA), which directly models the covariance of all features without independent assumption.

Students were tasked to label each scenario in the testing dataset. Their submissions were automatically evaluated by our script to report TPs, FPs, TNs, and FNs. The best-performing team was awarded three points as a bonus in their final grades. The second-place team got two points, and the third-place team got one point. All teams were evaluated based on the code quality, report quality, and classifier performance as their mid-term grade. The results from the top team are shown in Table \ref{tab:mid-term}.

\begin{table}[htbp]
    \centering
    \caption{Mid-term project results from the top three teams.}
    \begin{tabularx}{\linewidth}{
    >{\centering\arraybackslash}X
    >{\centering\arraybackslash}c
    >{\centering\arraybackslash}c
    >{\centering\arraybackslash}c}
        \toprule
        %\midrule
        Team & Metric & Quantity (IGA) & Quantity (MGA) \\
        %\midrule
        \midrule
        \multirow{4}{*}{Top-1} & TP & 38 & 36 \\
        & FP & 0 & 1\\
        & TN & 19 & 18\\
        & FN & 1 & 3\\
        % \midrule
        % \multirow{4}{*}{Top-2} & TP & 39 & 38 \\
        % & FP & 19 & 0\\
        % & TN & 0 & 19\\
        % & FN & 0 & 1\\
        % \midrule
        % \multirow{4}{*}{Top-3} & TP & 38 & 34 \\
        % & FP & 1 & 10\\
        % & TN & 18 & 9\\
        % & FN & 1 & 5\\
        \bottomrule
    \end{tabularx}
    \label{tab:mid-term}
\end{table}

\begin{figure}[htbp]
    \centering
    \includegraphics[width=0.8\linewidth]{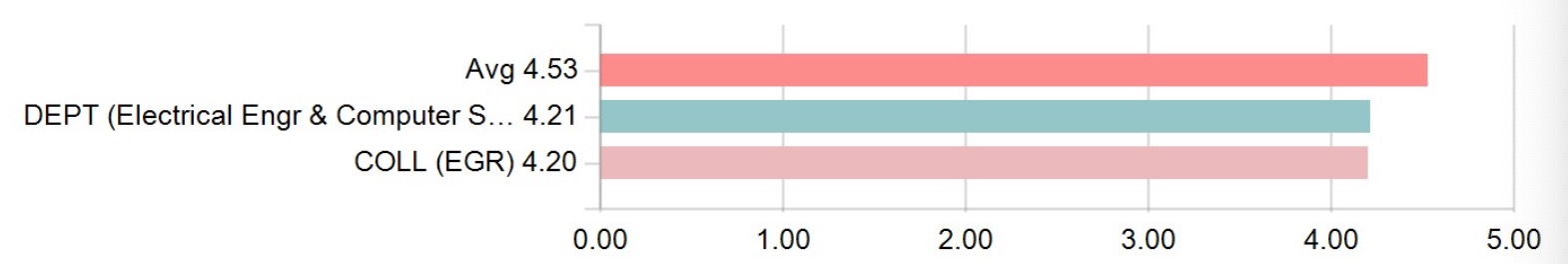}
    \caption{Course evaluation regarding real-world engineering perspective} 
    \label{fig:eval-1}
\end{figure}

\begin{figure}[htbp]
    \centering
    \includegraphics[width=0.8\linewidth]{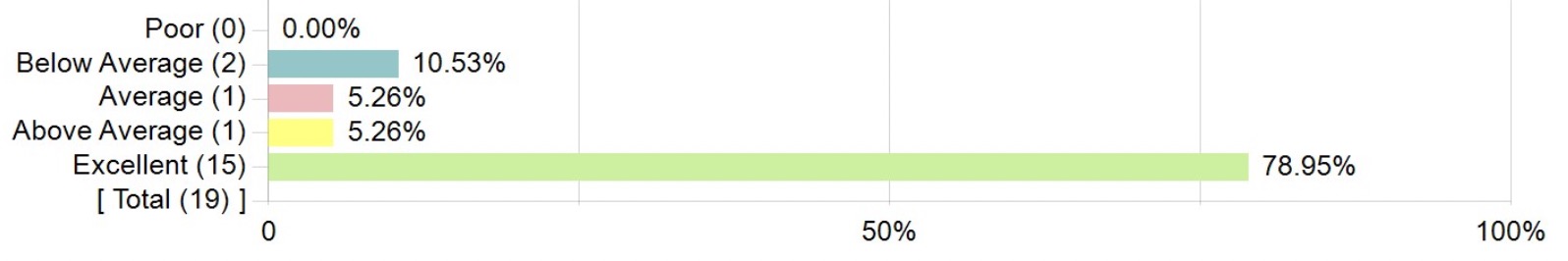}
    \caption{Evaluation statistics regarding real-world engineering perspective}
    \label{fig:eval-2}
\end{figure}

In terms of the real-world engineering perspective in the course evaluation, our course got 4.53 out of 5, which was significantly higher than the average at the department level and the college level (see Fig. \ref{fig:eval-1}). The detailed statistics in Fig. \ref{fig:eval-2} show that most of the students agree that they received real-world practice from our course.

%\vspace{-0.5cm}
\section{Conclusion and Future Work}
\label{sec:conclusion}
In this work, we first provide a reproducible framework for research and education in cybersecurity and ICS on two simulated platforms: GRFICS and Factory I/O. We implement an automated framework for launching cyberattacks, collecting data, and deploying machine learning-based intrusion detectors. This expands the capacity of the two simulations to allow future researchers to reproduce our experiment as well as design and compare their own intrusion detection models. In future work, we aim to use PLCs to create a hardware-in-the-loop setup for testing.

\section{Acknowledgment}
This material is based upon work supported by the National Science Foundation under Grants No. 2028397 and No. 2301543. This work is supported by the US Department of Education (Award\# P116S220004). This work is also in part supported by Ohio Cyber Range Institute Regional Programming Center.

\bibliographystyle{ieeetr}
\bibliography{ref}
\end{document}